# Conceptual Modeling Founded on the Stoic Ontology: Reality with Dynamic Existence and Static Subsistence

Sabah Al-Fedaghi[*]

*Computer Engineering Department*
*Kuwait University*
*Kuwait*

salfedaghi@yahoo.com, sabah.alfedaghi@ku.edu.kw

*Abstract* **– According to the software engineering community, the acknowledgement is growing that a theory of software development is needed to integrate the currently myriad popular methodologies, some of which are based on opposing perspectives. Conceptual modeling (CM) can contribute to such a theory. CM defines fundamental concepts to create representations of reality to achieve ontologically sound software behavior that is characterized by truthfulness to reality and conceptual clarity. In this context, CM is founded on theories about the world that serve to represent a given domain. Ontologies have made their way into CM as tools in requirements analysis, implementation specification, and software architecture. Currently, the primary role of ontology has been its use as a domain knowledge source rather than its classical concerns about** *being* **and** *existence***. However, modeling succeeds by facilitating and representing reality. We achieve a firm understanding of the modeled domain though comprehending the being and existence of its elements. This paper involves building a direct connection between reality and CM by establishing mapping between** *reality* **and modeling** *thinging machines* **(TMs). Specifically, Stoic ontology serves to define the existence of TM things and actions in reality. Such a development would benefit CM in addition to demonstrating that classical concepts in philosophy can be applied to modern fields of study. The TM model includes static and dynamic specifications. The dynamic level involves time-based events that can be mapped to reality. The problem concerns the nature of a time-less static description, which provides** *regions* **where the actions in events take place; without them, the dynamic description collapses. The Stoics came up with a brilliant move: the assumed reality to be a broader category than** *being***. Reality is made of things that** *exist* **and things that** *subsist***. This idea retains the commonsensical notion that both objects and concepts are in some sense** *real***. In this case, the dynamic TM description is in existence, whereas the static, mapped portion of the dynamic description is in subsistence. We apply such ontology to a contract workflow example. The result seems to open a new avenue of CM that may enhance the theoretical foundation for software and system development.**

---

## I. INTRODUCTION

According to Kirk and MacDonell [1], acknowledgement within the software engineering community is growing that a theory of software development is needed to integrate the myriad currently popular methodologies, some of which are based on opposing perspectives. Philosophical foundations and premises in software engineering are not yet well understood [2]. In recent times, an exploration of such foundations has occurred in search of a deeper understanding of the discipline [2]. This article aims to apply the classical Stoic ontology to the conceptual model, called *thinging machine* (TM), which has been developed to represent reality at a high level of abstraction. According to Dieste et al. [3], "Conceptual modeling is a crucial software development activity for both software engineering and knowledge engineering."

One of the implications of the word *engineering* indicates an approach that is theory-approved [4]. Some experts opine that software is built in a primitive way and that the role of modeling as a treatment for the weakness of software engineering has become more important [4]. In this context, conceptual modeling (CM) can contribute to such a theory of software development through defining fundamental concepts to create representations of reality and achieve ontologically sound software behavior that is characterized by truthfulness to reality and conceptual clarity [1]. Conceptual models should be based on theories about the world using *ontology* as a foundation, which builds on the assumption that ontology can help to improve the understanding of the way the world is constituted [5]. The term *ontology* has become a key word within modern computer science. According to Wand and Weber [6], the state of "our Information Systems will be only as good as our ontologies." Ontology in CM is the key focus of this paper to describe systems and develop a better understanding to capture and represent "conceptual schemas" [6] of domains [7].

Wand and Weber [6] pointed out that some ontologies created in computer science "are not always rooted in a sound foundation of more fundamental constructs like things and properties." According to Fonseca [8], philosophical study of ontology may help us find ways to build good ontologies. Evermann and Wand [9] noted the calls in computer science ontology literature for returning to the philosophical meaning of ontology. However, philosophy has different branches that have different assumptions about the world and the ways we understand it. As introduced by Greek philosophers, ontology is concerned with reality and the being and becoming of reality, its possibility and its actuality, its modes of existence and behavior, and the categories of reality, such as substance, space and time, relation, causality, and purpose [10]. According to Coffey [10], "In the domain of Ontology there are many scholastic theories and discussions which are commonly regarded as … merely historical."

The primary role of ontology within software engineering has been its use as a domain knowledge source for human interpretation and understanding. However, its classical concerns are claims about the nature of being, where *being* also has an abstract sense when we speak of "the being or reality of things." It may be used in "a collective sense to indicate the sum total of all that is or can be—all reality" [10].

### A. Paper Subject: Like Bunge's Ontology

The current dominant approach to describing a software system is the use of object-oriented techniques, for which several software modeling methods and languages have been proposed such as the object-oriented Unified Modeling Language (UML). Despite its origins in programming, UML has been extended for use as a conceptual modeling language to represent the involved domain. The suitability of UML for modeling "real-world" phenomena has been questioned. It is proposed to assign real-world semantics to UML constructs by mapping these constructs to the ontology of Mario Bunge [11]. Bunge's ontology views the world as made of things that possess properties. Properties are known via attributes, which are characteristics people assigned to things [12]. For Bunge, ontology must recognize, analyze, and interrelate concepts to produce a unified picture of reality [13].

In this paper, we map real-world semantics to TM modeling, which are then used to build conceptual models. Reality is represented based on Stoic ontology. In the present study, we introduce significant aspects of importance in such ontology. Such a project would benefit CM in addition to demonstrating that classical concepts in philosophy can be applied to modern fields of study.

### B. Illustration: TM Modeling with Stoic Ontology

Even though the TM model (sample recent paper [14] and old paper [15]) is still to be reviewed in the next section, in this subsection we need a simple example that illustrates the need for ontology in such a model. According to Hayes [16], much has been written about various views on how best to describe time and change. Several schools of thought have emerged, including the use of temporal modalities, a basic ontological distinction between things and processes and the situation calculus. These views seem to be inconsistent with one another, and the ontological frameworks derived from them are usually thought of as incompatible. Hayes [16] gave a simple assertion involving time, such as, *Bill was taller than his father and his aunt last year*, and considered ways to represent the content of such an assertion.

Hayes [16] presented several solutions, including the following two logical descriptions.

1. *(Taller(Bill, father(Bill)) & Taller(Bill, aunt(Bill))) AT t in LastYear*

This can be understood as meaning, "At any time in last year, it would have been true to say that Bill was taller than his father and his aunt."

2. *For all t in LastYear, Taller(Bill, father(Bill), t) & Taller(Bill, aunt(Bill), t)*

The situation assumes that the "things" in the situation being described are unchanging (i.e., they are continuants), but may have changing properties.

Because our purpose is to explain the problem through a modest example, we will simplify, without loss of generality, the assertion *Bill was taller than his father and his aunt last year* <u>to</u> *Bill was taller than his father last year*. Fig. 1 shows the static representation of *Some person is taller than another person*. The *create* action of persons 1 and 2 (pink numbers 1 and 2) indicates the two persons potentially *exist*. These persons have heights (3 and 4). Each height has a value (5 and 6). The relationship *taller* (7) exists between the heights of the two persons. Taller takes the values and processes them (8, 9 and 10); that is, compares them to create the result of comparison, say, value 1 > value 2.

To construct the dynamic model, we need to introduce the notion of a TM event. Like objects, events may be viewed as just the "content" of some portion of space-time [17]. Kim [18] identified an event as a triple (object, property, and time). Events in TM modeling involve injecting time in a region of the static model. Fig. 2 shows the event *The two values are compared*. Note that the region in Fig. 2 is a subdiagram of the diagram in Fig. 1. For simplicity sake, all events will be represented by their regions.

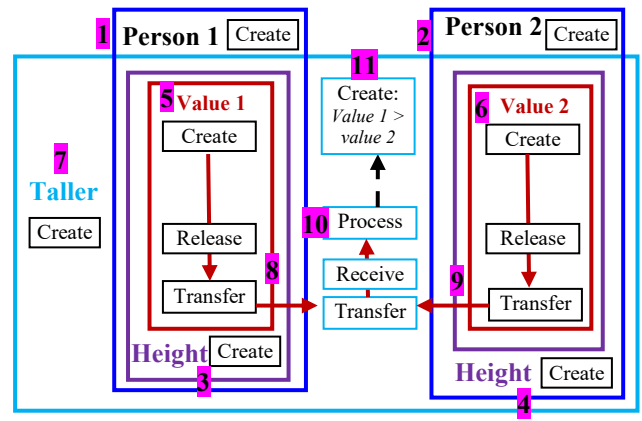

Fig. 1 Static TM model of *a person being taller than another*.

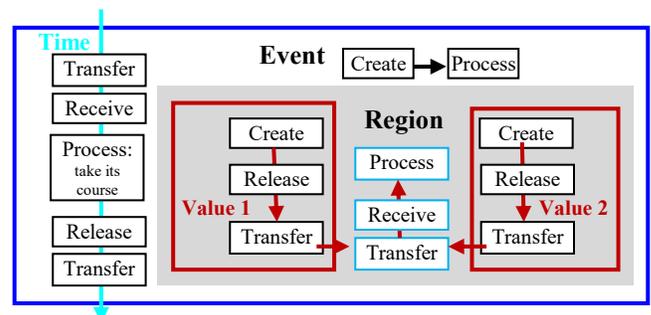

Fig. 2 The event *Two values are compared*.

Accordingly, Fig. 3 shows the following selected events.

$E_1$: Bill exists.

$E_2$: Bill's father exists.

$E_3$: Bill is described in terms of his height.

$E_4$: Bill's father is described in terms of his height.

$E_5$: Bill's height value is known.

$E_6$: The father's height value is known

$E_7$: The two known values are compared.

$E_8$: The comparison resulted in Bill's height > Bill's father height.

$E_9$: The relationship between Bill and his father has been established.

Fig. 4 shows the sequence of these events, called the behavior model. Figs. 3 and 4 express that *There exist Bill and his father who have height measurements that indicate that Bill is taller than his father*. All events are assumed to have occurred *last year* in the indicated chronology.

Note that the static model in Fig. 1 does not display entities (e.g., John) persistence or events (e.g., comparison) appearance in time. Fig. 4 shows the entities and events in their order of appearance and extensions (non-specious thickness), where "the past and future subsist, the present 'exists'" [19]. First, John's father exists and his height and its value appear, and then John and his attributes appear. Next, a comparison is executed to recognize the relationship. Time slots in the figure are not necessarily equal. As will be discussed later, entities and their attributes are viewed as extended things in the TM dynamic model.

Fig. 5 shows the system of static and dynamic TM levels. The static level includes all *regions* (of events, past, present, and future). In classical philosophical language, it embeds all past, present, and future *regions* in a branching structure of alternative possibilities. The dynamic level includes the progress of the execution of *events* in terms of sequence of *presents* (as referred to in the structure of metaphysical views of time). For example, present 1 requires that $E_1, \ldots, E_6$ be currently realizable (existing) for present 2 to become reality. The blue lines in Fig. 5 indicate extended events (e.g., John appears in present 1 and exists in present 2, present 3, present 4, etc. Fig. 6 illustrates the timing of different events.

As illustrated in this example, TM modeling involves two levels that are characterized by staticity and dynamics. *Staticity* refers to a static model that represents the world of potentialities outside time where everything is present simultaneously. According to Finton [20], "Potentiality is not something that exists. It has no definite time and occupies no space. In this case, potentiality is the invisible seed of something that can possibly exist."

The dynamic model represents the world of actuality in time where events do not necessarily happen simultaneously. Time is a thing that creates, processes, releases, transfers, and/or receives other things. The two TM models share regions (sub-thinging machines). TM staticity and dynamics are applied not only to things but also to actions, thus we have static (potential) actions in the static TM model and events (actual actions) in the dynamic TM model.

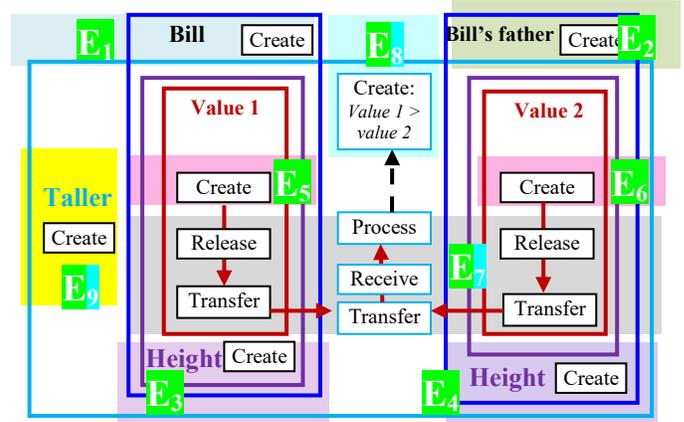

Fig. 3 Static TM model of *Bill is taller than his father*.

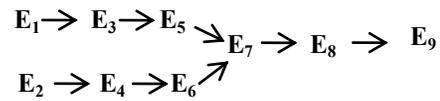

Fig. 4 The behavior model.

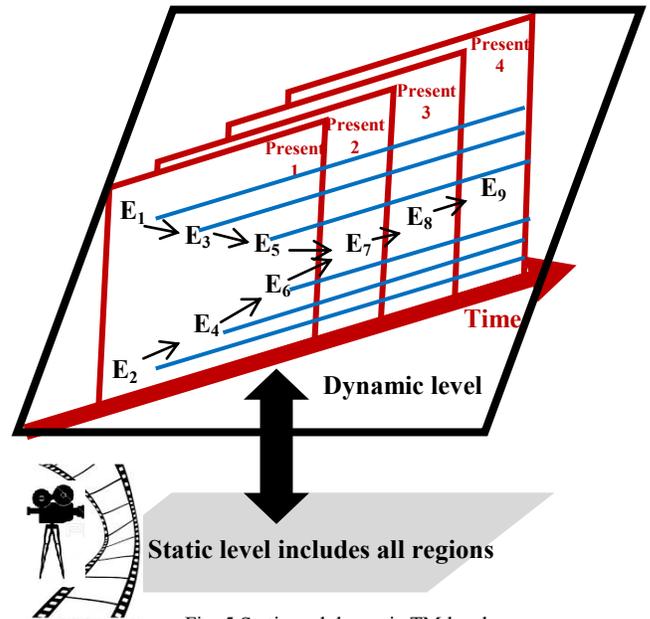

Fig. 5 Static and dynamic TM levels.

| Time slot | T1 | T4 | T6 | T1 | T3 | T5 | | | |
|-----------|----|----|----|----|----|----|---|---|---|
| $E_1$ | John | | | | | | | | |
| $E_2$ | Father | | | | | | | | |
| $E_3$ | Height-J | | | | | | | | |
| $E_4$ | Height-F | | | | | | | | |
| $E_5$ | Value-J | | | | | | | | |
| $E_6$ | Value-F | | | | | | | | |
| $E_7$ | Compare | | | | | | | | |
| $E_8$ | Create | | | | | | | | |
| $E_9$ | Taller | | | | | | | | |

Fig. 6 Events sequence and persistence in time.

TM staticity clearly includes universal TM things that describe common features or relationships, Plato's "ideas" or "forms," sets, numbers and other static categories [21]. A static model displays an atemporal diagrammatic description. Decomposing the static model into subdiagrams forms the regions of temporal events.

With this introductory understanding of TM modeling, we can state the problem of concern in this paper.

### C. Problem: Fitting the TM Model to Reality

Modeling has its roots in the study of real-world problems. Thus, using a conceptual model in such a context requires an ontology that provides a framework for the way problems should be understood and addressed, as well as the way to define the "reality of the model" (i.e., how the model fits reality). The definition of reality assists in deciding how to find something to which the model can be applied in reality. In modeling, ontology concerns "articulating the nature and structure of the world" [22].

However, how do we achieve a firm understanding of the modeled domain without comprehending the being and existence of its elements (e.g., entities and relationships)? This paper aims to build a direct bond between reality and conceptual models by defining a way to map reality using the TM model.

Note that the crucial analysis step of TM modeling involves the move from staticity to dynamism (e.g., Fig. 1 and Fig. 3 in the previous example). The dynamic level involves time-based TM events that can be mapped to reality. We can claim that the dynamic model reflects reality (i.e., reality can be described by this model). This claim of modeling reality provides a "sense of reality." The problem is concerned with the reality of the time-less (conceptual) static description that provides the events' regions; without them, the dynamic description collapses.

### D. Solution: Stoic Ontology

It has been suggested that the Stoic ontology should be conceived as a reaction against Platonism. Platonism is the view that an abstract object that does not exist in space or time and is therefore entirely nonphysical and nonmental [23]. The central claim of Stoic ontology is that only bodies exist [24]. Plato assumes that for something to be a thing, it must have "being" (i.e., existence). He suggests that existence is marked by the capacity to act or be acted upon. Stoics must either admit that concepts do not exist or that they are made of matter. Stoics agree that everything that exists is made of stuff and can act or be acted upon. [25].

> And here comes the stroke of genius: reality is assumed to be a broader category than being. That broader category is made of two sub-categories: things that exist and things that "subsist." This is a brilliant move, because it retains the commonsensical notion that both physical objects and concepts are in some sense "real," but makes explicit exactly what this means: bodies are real and they exist, concepts are real and they subsist.

Which explains why bodies, but not concepts, are made of stuff and can act or be acted upon. [25]

Accordingly, reality consists of both existing and subsisting things.

In this paper, we propose that TM modeling reflects reality where the dynamic model specifies what *exist* and that the part of the static model mappable to the dynamic level describes what *subsists*. Time "translates" or maps static things and actions *into* actuality and the non-mappable things in the static level designate things that do not exist or subsist, such as unicorns and "Plato's transcendent forms" [26]. The static model mappable to the dynamic level will be called in this paper the *mappable* model, description, or specification.

Consider the assertion *dinosaurs exist*, which is a topic in a philosophical controversy that denies and affirms it [27]: Those who endorse the claim understand the verb *exist* to be tenseless, and those who deny it take *exist* to be a present tense verb. Fig. 7 shows the TM modeling solution based on the Stoic ontology where *reality* is assumed to consist of two subcategories: things that exist and things that subsist. Fig. 8 shows a non-mappable thing that neither subsists nor exists.

### E. Paper Structure

This paper is structured as follows. The next section gives a brief description of the TM model with a new contribution. Section 3 is a sample of use of ontologies to provide a blueprint of the space to capture semantics from domain. The section includes a proposed contract workflow model using a contract ontology knowledge base. In Section 4, we explore some of the notions in the Stoic ontology, including traffic as a subsisting thing and *lekta* (what gets said).

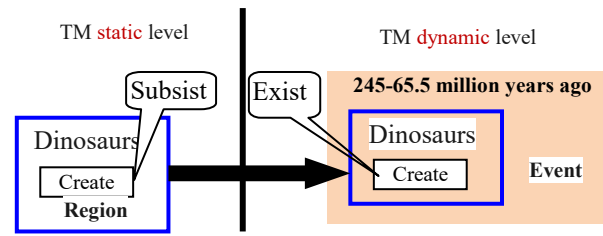

Fig. 7 In Stoic ontology, *reality* is assumed to consist of two subcategories: things that exist and things that subsist.

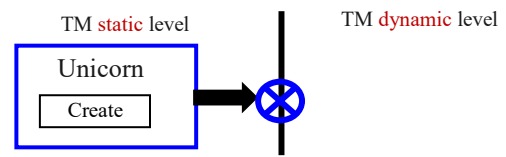

Fig. 8 The third thing in the Stoic ontology neither subsists nor exists.

## II. THING MACHINE (TM) MODEL

In the TM model, entities are conceptualized as thimacs (*thi*ngs/*ma*chines). A thimac represents a type of "state of affairs," similar to Minsky's frames [28]. As mentioned in the previous section, the TM modeling has two levels of specification:

-   A static model that represents static things and static actions. The five specific static actions are create, process, release, transfer, and release.
-   A dynamic model that includes a static model subdiagram (called a *region*) and time, leading to the construction of events (i.e., the realization of static entities and actions).

Time is an important element in such a view. Note that time is incorporeal in Stoic physics. Only existing things (corporeal things) can causally interact as agents or patients. According to Greene [29], Stoic time is "a *dimension* of bodies that contains (but is not identified with or a property of) the *motions* of bodies. Time is still a something, despite not being a corporeal something [italics added]." In TM modeling, *staticity* refers to a static model that represents the world of potentialities outside time where everything is present simultaneously. The dynamic model represents the world of actuality in time where events do not necessarily happen simultaneously. Time is a thimac (i.e., a thing that creates, processes, releases, transfers, and/or receives).

Only thimacs that can embed time are realizable (exist) at the dynamic level. Thus, for example, a "square circle" is a static thimac that cannot be injected with time to exist in the dynamic model, and neither can it subsist because it is not mappable to the dynamic level. Static thimacs have unique names; if two static thimacs have the same name, then they are injected with different time slots to be realizable at the dynamic level. Contradictory thimacs (regions) occupy the static level, described in philosophy as an "inconsistent multiplicity," but only consistent regions are realized at the TM dynamic level. Subsisting contradictories exist one at a time (e.g., if p and ¬p subsist, then if dynamic modeling is applied, either p or ¬p exists). The TM universe is populated by things that exist/subsist at one or two levels of being.

A thimac has a dual mode of being: the machine side and the thing side. The machine has the (potential) actions shown in Fig. 9. The sense of "machinery" originates in the TM actions indicating that everything that creates, processes, and moves (i.e., releases, transfers, and receives) other things is a machine. Simultaneously, what a machine creates, processes (changes), releases, transfers, and/or receives is a thing. In this view, the "world" is a totality of thimacs.

### A. *The Machine*

In Stoic ontology, a "body" is anything that can act or be acted upon. It is the mark of existence. This is a TM dynamic thimac. A thimac is a machine that creates, processes, releases, transfers, and/or receives; simultaneously, it is a thing that can be created, processed, released, transferred, and/or received.

TM actions shown in Fig. 9 can be described as follows.

**Arrive**: A thing moves to a machine. In Stoic ontology, motion exists along the temporal dimension and "the length of a motion – or its duration – can be measured by how much of this temporal dimension the motion covers. One motion has a longer duration than another just in case it covers more of the temporal dimension" [29]. In TM modeling, not only is time the dimension in which motion exists, but it is also the dimension in which actions (create, process, release, transfer, and receive) exist.

**Accept**: A thing enters the machine. For simplification, we assume that arriving things are accepted; therefore, we can combine the **arrive** and **accept** stages into the **receive**.

**Release**: A thing is ready for transfer outside the machine.

**Process**: A thing is changed, handled, and examined, but no new thing results.

**Transfer**: A thing is input into or output from a machine.

**Create**: A new thing "becoming" (is found/manifested) is realized from the moment it arises (emergence) in a thimac. Simultaneously, it also refers to the "existence" of a thing, especially where we want to emphasize persistence in time as illustrated next.

Additionally, the TM model includes a *triggering* mechanism (denoted by a dashed arrow in this article's figures), which initiates a (nonsequential) flow from one machine to another. Multiple machines can interact with each other through the movement of things or through triggering. Triggering is a transformation from the movement of one thing to the movement of an another thing. The TM "space" is a structure of thimacs that forms regions of events (to be defined later). Note that for simplicity's sake, we may omit *create* in some diagrams, assuming the box representing the thimac implies its existence (in the TM model).

### B. *Two Level of Representation and Lupascian Logic*

The TM model is different from similar notions that are currently used in the literature. Note that the TM approach takes the side of philosophers (e.g., Whitehead), who conceive of *objects* as things that extend across time. Objects and events are things of the same kind [30]. The TM model differs from such an ontological approach by introducing the notion of thimac as a thing and machine extended in time.

TM modeling is based on distinguishing two worlds: potentialities, described in terms of static actions (do not embed time), and actualities (dynamics), represented by (temporal) events. This implies the activation of an event and its *negative event* can be defined as an alteration between the static level and the dynamic level.

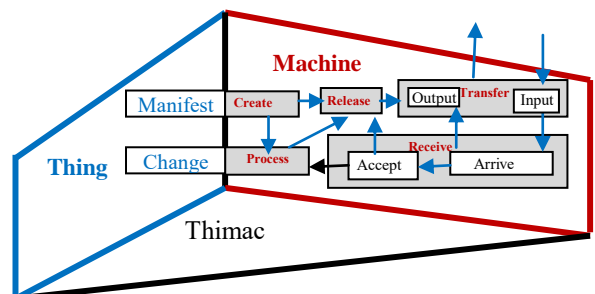

Fig. 9. The thimac as a thing and machine.

Instead of "process" versus "stop process," we have event (which includes the region to the dynamic level) versus "revert to static process," which returns the region of event to the static level. We take this method of eliminating negativity from the philosopher Stéphane Lupasco. According to Brenner [31], every element *e* (an event in TM modeling; i.e., a thimac that contains a region plus time) is always associated with a *non-e* (a static thimac in TM modeling), such that the actualization of one entails the potentialization of the other and vice versa, "without either ever disappearing completely" [31]. Additional illustration of this topic can be found in Al-Fedaghi [14]. The next section in this paper includes an example of negative events that are represented according to Lupascian logic.

In Lupascian logic, a negative event is a return to potentiality from being an event. After the event *telephone call* ends with the negative event *there is no telephone call (hanging up the telephone)*, the caller, the receiver of the call, the telephone, etc., still exist, but they have returned to the non-eventness state. negative events come in several types, but representing them in TM modeling is a topic for a future paper. Note that the Stoics' ontology allows continuous passage from corporeal to incorporeal and back again.

*C. Events*

According to Salles [32], Stoics would distinguish between two kinds of events: those consisting in "qualitative" exemplification and those consisting in a "dispositional" one. For example, *my coffee is hot* would be the exemplification of my coffee having the property of being hot, which is a qualitative state of my coffee because it is explained by the presence in my coffee of heat, which is a quality. Fig. 10 shows the corresponding static and dynamic models in TM modeling. Simply put, the figures indicate *I as an existing thing have coffee that is hot*. However, the event consisting of *my walking*, the exemplification by me having the property of walking, does not consist of a qualitative exemplification, but in a dispositional one because it is not explained by the presence in me of some supposed quality of walkingness; rather, it is simply explained by my being disposed in a certain way [32]. Fig. 11 shows the corresponding TM dynamic and behavior models. The static model can easily be extracted from the dynamic model. Accordingly, it seems that the specific notion of event is different from the Stoic notion of event. Such a topic needs to be further researched in the future.

### III. TM Example: Contract Workflow Model

According to Kabilan [7], systems rely upon ontologies to provide a blueprint of the space to capture semantics from domains, standard terms, and vocabularies. Kabilan [7] proposed a contract workflow model to visualize a contract as containing prescriptive directions for the execution of business activities or tasks that would in turn fulfill the obligations expressed in the contract.

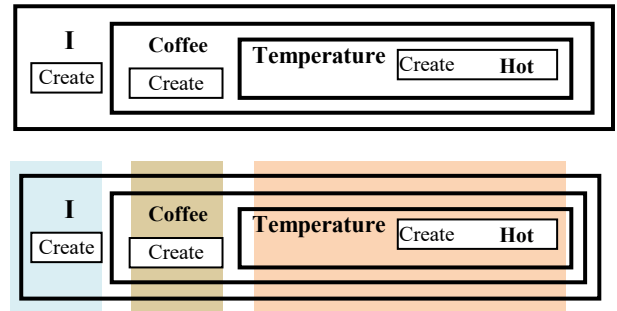

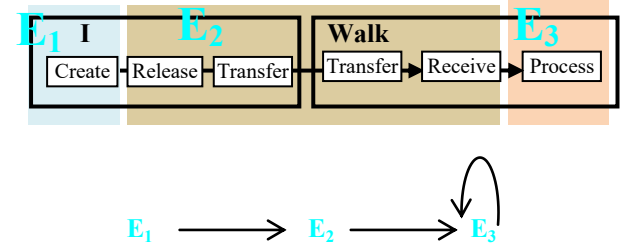

Fig. 10 TM models of *My coffee is hot*.

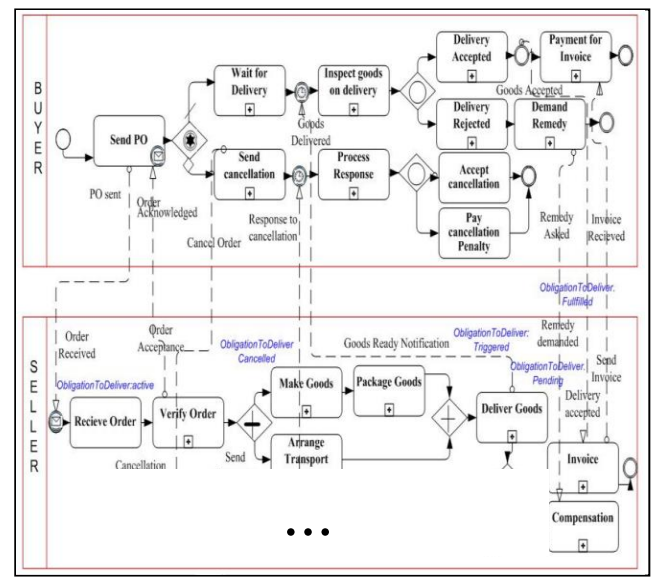

Fig. 11 TM dynamic and behaviour models of *I am walking*.

The model is derived from a given business contract instance, using a contract ontology knowledge base.

Using BPMN notations, the user is required to represent the performance of events as tasks, and the individual obligation-performance diagram as two business processes, using swim lanes. In the case of a secondary obligation, all the related activities may be modeled as subprocesses using the BPMN. Fig. 12 shows the resultant BPMN model.

Fig. 12 The BPMN model (from [7]).

Events are modeled as entity types, which, according to Kabilan [7], allow us to model behavioral aspects of a domain. Events are things that happen. Accordingly, Fig. 12 includes both the static and dynamic specification of the system.

Fig. 13 shows the static TM model. First, the buyer creates an order (pink number 1) and sends it to the seller (2 and 3) where it is processed (4). If the order is accepted (5) (Kabilan [7] did not include rejection, hence we do the same), then

- An acknowledgement is sent to the buyer (6).
- Acceptance triggers the process of creating the ordered product (7).
- Acceptance triggers the process of transport arrangement (8).

Eventually, the product (9) and transport arrangement (10) are sent to the delivery department to be processed (11). This process results in sending (12) the product to the buyer (13). The buyer processes (14) the product to decide whether to accept (25) or reject (16) it. Accepting the product (15) and receiving an invoice from the seller (17 and 18) trigger the buyer to send payment to the seller (19 and 20).

Note that the horizontal thick bar (21) is used to simplify the diagram by indicating that both conditions are necessary to send the payment. If the product is rejected (16), then a demand for remedy is sent to the seller (22 and 23). The demand triggers the seller to send compensation to the buyer (24 and 25). In case of cancelation (26 and 27), then upon receiving the cancelation, the seller's process is (28) as follows:

- The process of transport arrangement of the product is stopped (9). This is a negative action as indicated by the diamond-head arrow (30).
- A response to the cancelation is sent to the buyer (31 and 32). Accordingly, the buyer sends a cancelation fee (33 and 34).

Accordingly, the dynamic model is built using the following events as shown in Fig. 14.

$E_1$: The buyer creates an order and sends it the seller.
$E_2$: The seller receives the order and processes it.
$E_3$: The order is accepted.
$E_4$: An acknowledgment is sent to the buyer.
$E_5$: The seller makes transport arrangement.
$E_6$: The seller **creates the ordered product.**
$E_7$: The product and transport arrangement are received by the delivery department.
$E_8$: The product is sent to the buyer.
$E_9$: The buyer processes the product.
$E_{10}$: The buyer accepts the product.
$E_{11}$: The buyer rejects the product.
$E_{12}$: An invoice is sent to the buyer.
$E_{13}$: The buyer sends a payment.
$E_{14}$: The buyer demands a remedy.
$E_{15}$: The buyer receives compensation.
$E_{16}$: The buyer cancels the order.
$E_{17}$: The seller sends a cancelation response.
$E_{18}$: The buyer sends a payment for cancelation.

Additionally, there is the negative event R5: Stoppage of delivery arrangement. Here, we follow Kabilan's [7] description, which does not discuss what happens in case of cancelation.

Fig. 15 shows the behavior model.

## IV. EXPLORATION IN THE STOIC ONTOLOGY

The Stoics divide "what is" into three species: things that are corporeal (e.g., body, TM dynamic things), things that are incorporeal (TM mappable static things), and things that neither subsist nor exist [26].

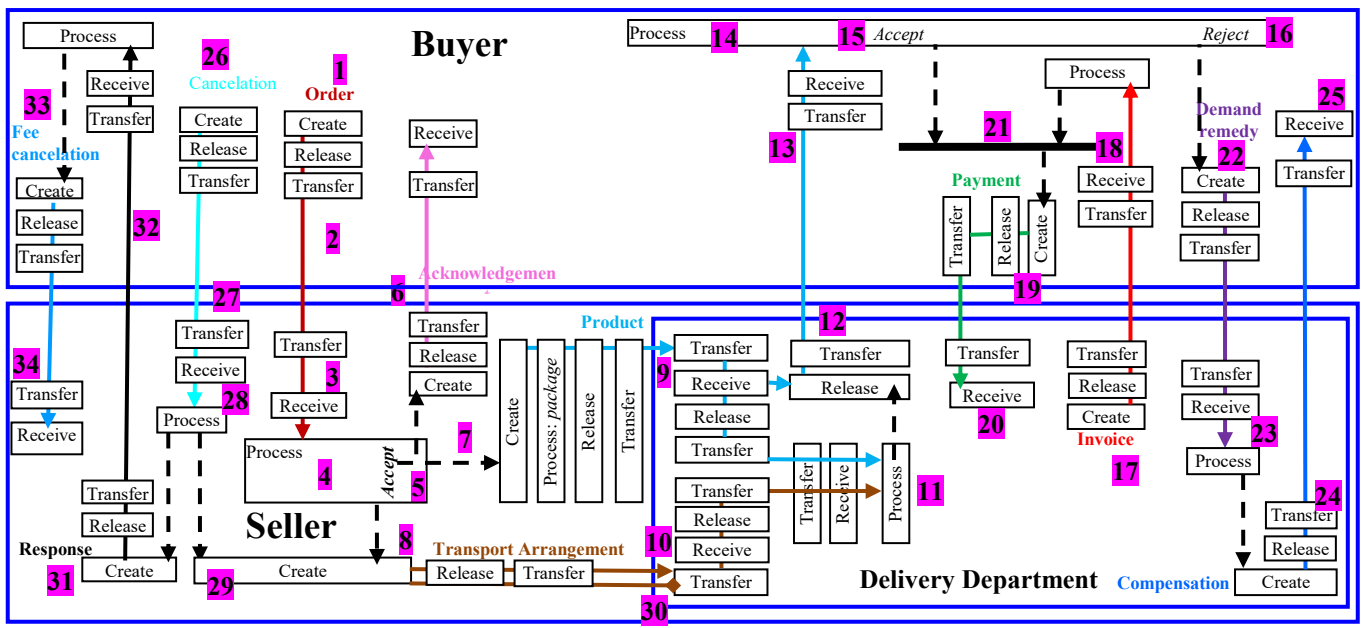

Fig. 13 Static model.

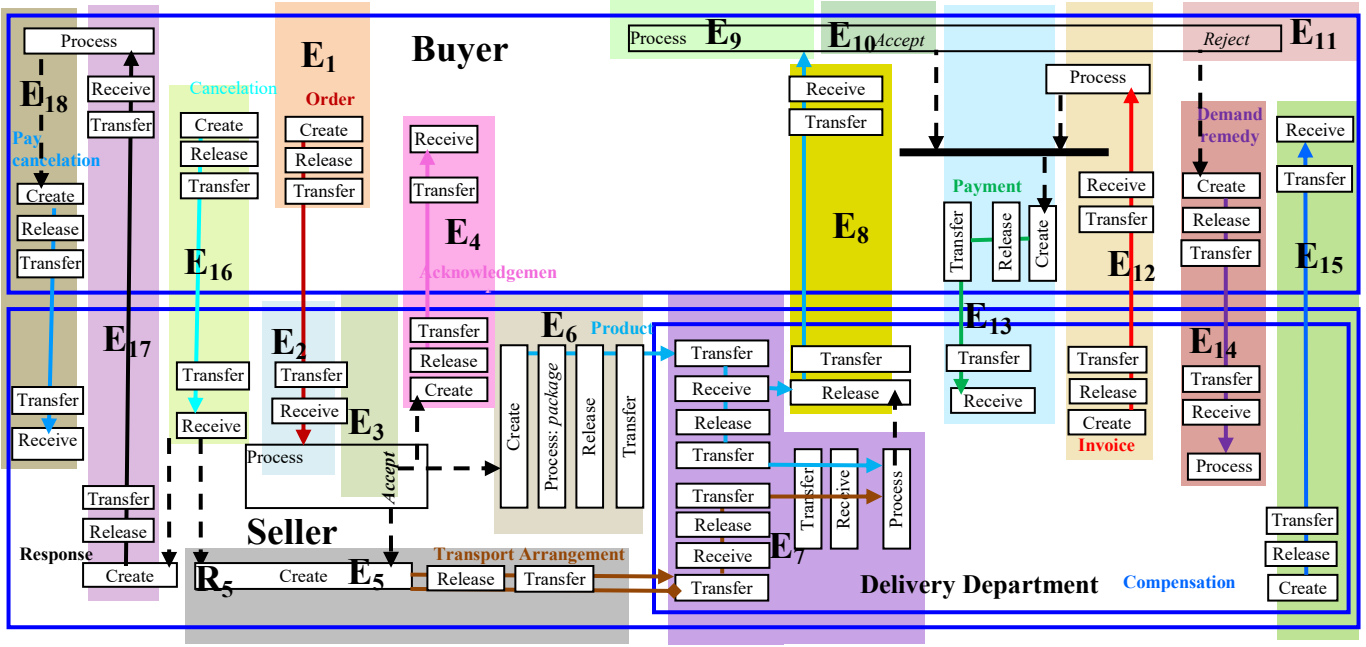

Fig. 14 The dynamic model.

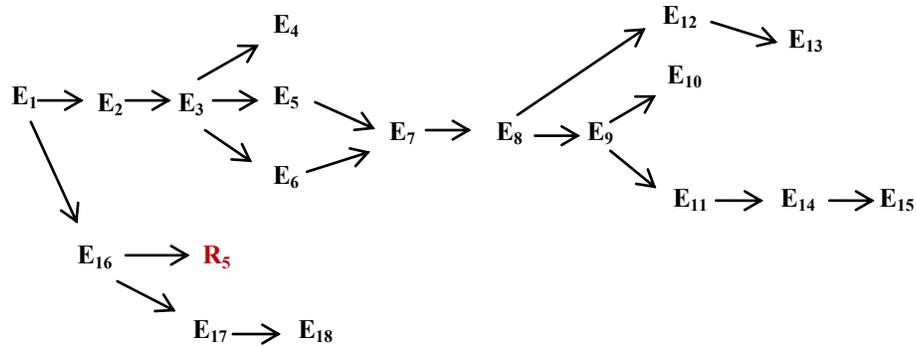

Fig. 15 The behavior model.

- All existent things are particulars, and time is the "dimension of motion" or "of the world's motion" [33]. Bodies are malleable and continuous [34]. These bodies are viewed as special types of events in the TM dynamic model.
- Immaterial entities such as concepts and mathematical (mappable) things exist as in a derivative mode of reality, call subsistence; these things are nonexistent in that they are not themselves solid bodies, but they are nonetheless something physical because they depend on bodies for their subsistence [34].
- The subsisting things will be viewed as the mappable part of the static TM model.

Incorporeal things (mappable things in the static description) depend on what is called "bodies without themselves being bodies," much as the flow of traffic depends on cars without being "nothing" but the cars [35].

Though these entities do not exist in the manner of bodies, they nonetheless have a derivative kind of reality the Stoics term *subsistence* [35]. Although they are not (physical) bodies, they are an ineliminable part of the world's objective structure [36].

### A. Example of Incorporeal Things

According to Harven [37],

> The incorporeals can each be understood as body-less: entities that depend on body without themselves being bodies, much like the flow of *traffic* depends on cars without being reducible to the cars that give rise to it. The flow of traffic is a proper object of thought and discourse—we can say true and false things about it, like that it is slow or fast; and it is objectively available for anyone to think about. In addition, such an entity will pass the Not-Someone test for particularity: the flow of traffic on the Bay Bridge cannot be the same flow of *traffic* as that on the Golden Gate Bridge. They might both be slow, stop-and-go, smooth, or fast; but just as my car cannot at the same time be on the Bay Bridge and the Golden Gate Bridge, so too the flow of traffic arising from cars on one bridge cannot be the same as that on another bridge. If a certain flow of traffic arises from cars in Athens, then it cannot be in Megara [italics added].

To illustrate traffic, we need to model it. Because it is impractical to model it due to space consideration, we will model a small part of a traffic system. Silvestre and Soares [38] modeled *waiting signal and pedestrian crossing* using a UML state machine as shown in Fig. 16. According to Silvestre and Soares [38], the state machine diagrams are useful in the real-time and embedded domain. They allow the description of a system behavior in terms of states and transitions between these states.

Fig. 17 shows a simplified static TM model of *waiting signal and pedestrian crossing*. The pedestrian (pink number 1) arrives at the sidewalk (2) to cross the street.

- If the pedestrian light is green (3), then the pedestrian crosses the street (4 and 5).
- If the pedestrian light is red (6), then the pedestrian waits (7) and presses the button to create a signal (8) that flows to the light control (9) to trigger the green light (3).

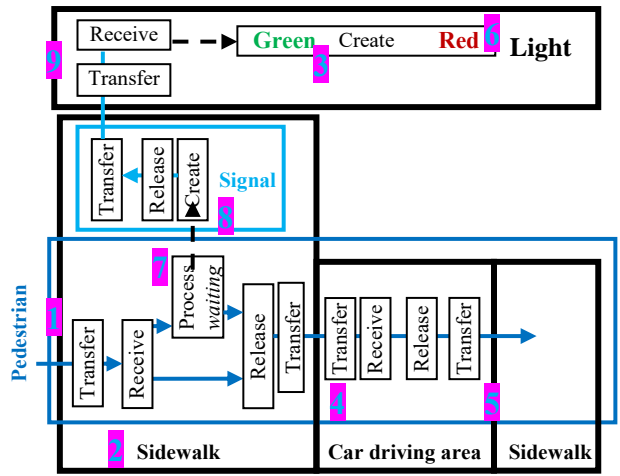

Fig. 17 TM static model.

Figs. 18 and 19 show the dynamic and behavior models, respectively. In Fig. 19, the big green and red ellipses denote "bigger" events that include the involved set E1 and E4 and set E2 and E4.

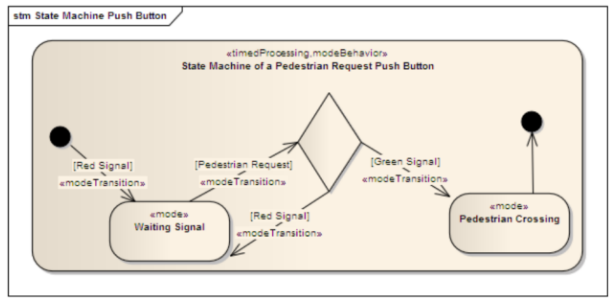

Fig. 16 UML state diagram of pedestrian crossing (from [38]).

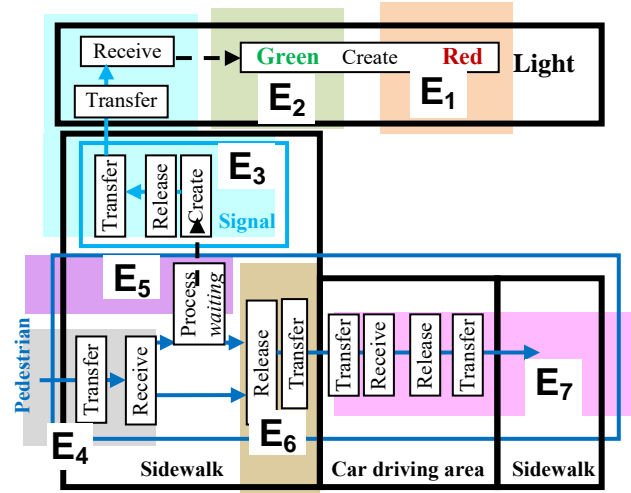

Fig. 18 TM dynamic model.

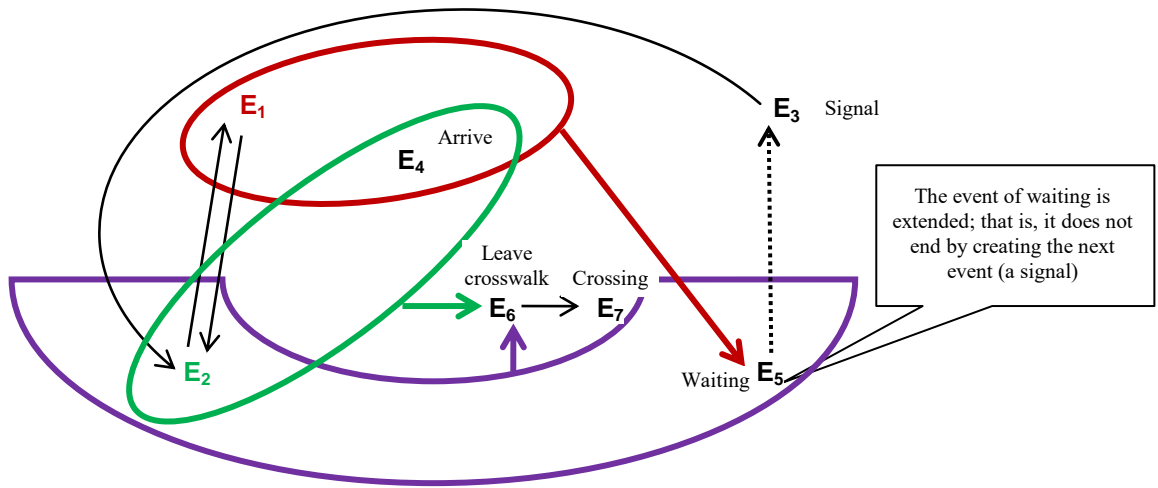

Fig. 19 TM behaviour model.

Note that the dotted arrow in Fig. 19 indicates that $E_5$ is to be kept "running" (extended) when transferring to the next event. In general, a nonentity TM event finishes when the next event begins, such as the events *Compare*, *Create*, and *taller* (Fig. 6) in the example of *Bill was taller than his father last year* in the introduction.

The point of this example is to model traffic (represented in this simple example by *waiting signal and pedestrian crossing*). We can claim that *traffic* demonstrates that an immaterial thing subsists. It is an incorporeal that is captured in the TM static model (Fig. 17). It *exists* at the dynamic level (Fig. 18) as concrete processes in terms of things and actions.

### B. Example 2 of Incorporeals: Lekta

The Stoics postulate four types of incorporeals: *sayable* (the meanings expressed by acts of speech, which we would call *concepts* [25]), *void* (empty space), *place*, and *time* [26]. For the Stoics, only bodies can act and be acted upon: In TM modeling, actions subsist as static actions in a TM or exist as actual action at the dynamic level. The incorporeals include potential TM actions. Subsistence is neither material being nor nothing, but somewhere between both [26].

The Stoics considered the sayable, (i.e., *lekta*, or what gets said) as one of the four incorporeals that lack what bodies have, namely the capacity for agency or of being passive receptors of an action [39]. What we say, or what gets said, is a *lekton* (the sense or meaning of a linguistic proposition) that is different from the involved utterance or sounds. What gets said is an incorporeal thing (independently present in reality, regardless of the sentences we utter or not), but sounds are corporeal. Uttering and saying are not identical.

Applying the concept to TM modeling, the sound of uttering is the action of issuing sound as a corporeal event at the dynamic level, whereas what is said is the lekta of the incorporeal region of utterance that becomes sound. Consider, again, Hayes's [16] simplified assertion *Bill was taller than his father*, which was discussed in the introduction, as the sayable. In the context of an utterance, it is modeled as shown in Fig. 20. In the figure, the lekta includes the utterance's static meaning. Fig. 21 shows the dynamic model. The event sound *Bill is taller than his father* being said expresses the dynamic existence of lekta (i.e., the actual meaning).

### V. CONCLUSION

In this paper, we have explored the application of the Stoic ontology to the TM model. A first impression is that the Stoic ontology fits hand-in-glove with the TM model. In this context, the static TM model seems to provide a portrait of "grounds of being" or "reality before existence." Such an initial observation is worth further research. This also needs further understanding of differences in assumptions and the ways the Stoic ideas diverge from TM modeling. The topic also needs scrutiny and better knowledge in Stoic ontology to determine the worthiness of this connection with conceptual modeling.

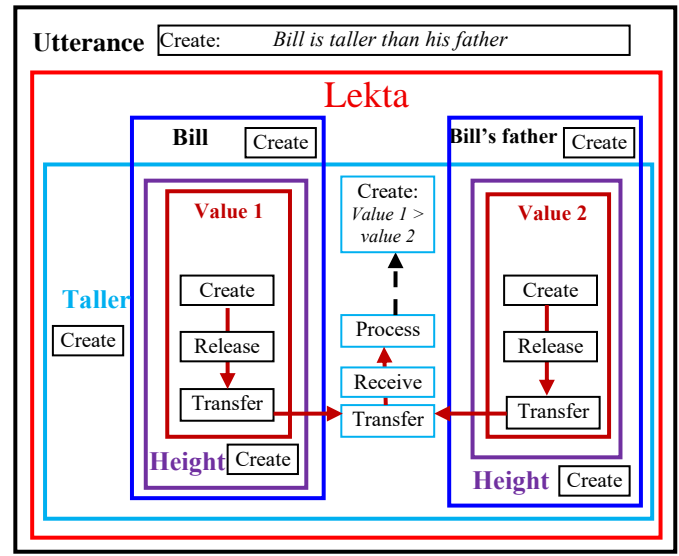

Fig. 20 The lekta of *Bill is taller than his father*.

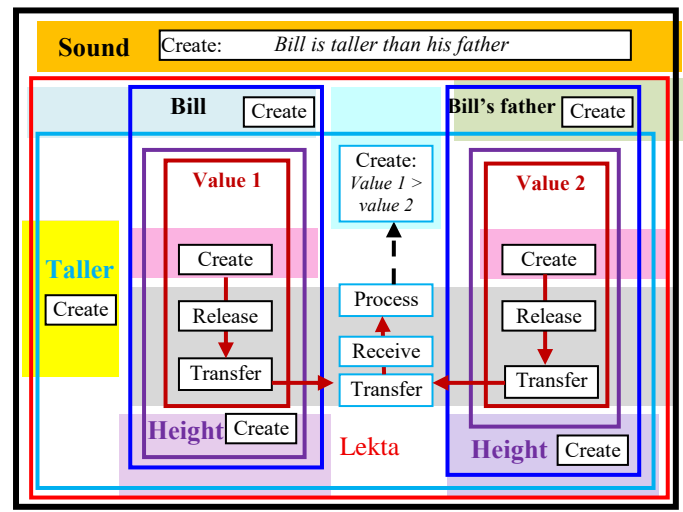

Fig. 21 Dynamic TM model of *Bill is taller than his father*.